\documentstyle[aps,prl,twocolumn,epsfig,amssymb]{revtex}
\def\B.#1{{\bbox{#1}}}
\def\x{{\mbox{\boldmath$x$}}}

\def\r{{\mbox{\boldmath$r$}}}

\def\unitr{{\mbox{\boldmath$\hat r$}}}

\def\f{{\mbox{\boldmath$f$}}}

\def\k{{\mbox{\boldmath$k$}}}

\def\v{{\mbox{\boldmath$v$}}}
\def\unitr{{\mbox{\boldmath$\hat r$}}}

\def\la{{\langle}}
\def\ra{{\rangle}}

\newcommand{\be}{\begin{equation}}
\newcommand{\ee}{\end{equation}}
\newcommand{\bea}{\begin{eqnarray}}
\newcommand{\eea}{\end{eqnarray}}

\begin{document}
\twocolumn[\hsize\textwidth\columnwidth\hsize\csname
@twocolumnfalse\endcsname
\title{Anisotropic Homogeneous Turbulence: hierarchy and intermittency of scaling exponents in the  anisotropic sectors.}
\author{Luca Biferale$^1$ and  Federico Toschi$^{1,2}$}
\address{$^1$Dipartimento di Fisica and INFM , Universit\`a "Tor Vergata",
Via della Ricerca Scientifica 1, I-00133 Roma, Italy}
\address{$^2$  University of Twente, Department of Applied Physics, Enschede, The Netherlands}

\maketitle
\begin{abstract}
We present the first  measurements of
anisotropic statistical fluctuations in {\it perfectly}
homogeneous turbulent flows. We address both problems of intermittency
 in anisotropic sectors and hierarchical ordering of  anisotropies
on a direct numerical simulation of a three dimensional {\it random} 
Kolmogorov flow. 
We achieved an homogeneous and anisotropic statistical ensemble
by randomly shifting  the forcing phases. 
We observe high intermittency as a function of the order of the velocity
correlation within each fixed anisotropic sector and a hierarchical
organization of scaling exponents at fixed order of the velocity 
correlation at changing the anisotropic sector. 
\vskip0.2cm
\end{abstract}

\vskip0.3cm
\vskip0.2cm
]

\noindent
At the basis of  Kolmogorov  1941 theory
there is the idea of  restoring of universality and isotropy at small scales
in turbulent flows. Memory of large scale anisotropic
forcing and/or boundary conditions should be quickly  lost during the process
of energy transfer toward small scales. The overall result being a local
recovering of isotropy and universality for turbulent fluctuations
at  small enough scales and large enough Reynolds numbers.\\

In recent years, a quantitative investigation of restoring of isotropy
in experimental anisotropic turbulence\cite{gw98,war00},
 numerical homogeneous shear flows
\cite{p96,ps96} and numerical channel flows \cite{bv00} 
questioned the main Kolmogorov paradigm, speaking explicitly
of {\it persistence of anisotropies}.
Some  theoretical work has also  been done \cite{alp00} 
in order to understand how to properly
link the invariance under 
rotation (SO(3) symmetry group) of the Navier-Stokes equations and the
analysis of anisotropic fluctuations of velocity turbulence correlations. \\
The observed  anisotropic effects in small scales
turbulence is both  a {\it theoretical } challenge and a very actual
{\it practical} problem, opening the question whether any, realistic,
anisotropic turbulent flows can ever possess  statistical features
independent of the (anisotropic) boundary and forcing effects. 
This goes under the name  of universality.\\
Neglected  anisotropic effects in
 high Reynolds numbers flows have also been 
proposed to be at the origin of  different statistical 
properties measured for  transversal and longitudinal velocity fluctuations
\cite{sreene_t}. 
Importance of  properly disentangling isotropic and anisotropic fluctuations
has also been demonstrated in the analysis of intermittency in channel
flow turbulence \cite{abmp99}.\\
Important step forward in the analysis of anisotropic
fluctuations have recently been done  
in Kraichnan models, i.e.
 passive scalars/vectors 
 advected by isotropic, Gaussian
and white-in-time velocity field with large scale anisotropic forcing
\cite{kra94,gk95,v96,lm99,abp00,arad_pass}.\\ In those models, 
anomalous scaling arises as the results
of a non-trivial null-space structure of the advecting
operator. In these  cases,  correlation functions in different sectors of
the rotational group show different scaling properties. Scaling
exponents are universal: they do not depend on 
the actual value of forcing and boundary conditions, and they 
are fully characterized by the order of the anisotropy. Non universal
effects are felt only in  coefficients multiplying the 
power laws. Coefficients  are fixed, in principle,  by requiring matching 
with  non-universal boundaries conditions
in the large-scale region. \\ Similar problems, like the very 
 existence of scaling laws in the anisotropic
sectors and, if any, what are  the values of the scaling exponents and 
what is the  dependency from universal/non-universal effects are at the 
forefront of experimental, numerical and theoretical research in true
turbulent flows. Only a few indirect experimental 
investigation of scaling in different
sectors \cite{arad_exp,sk00} and direct
 decomposition in Channel flow simulations \cite{abmp99,blmt00,bv00} have, at
the moment, been attempted. \\ The situations is still unclear, evidences
of a clear improving of scaling laws by isolating the isotropic sector
have been reported, supporting the idea that the undecomposed 
correlations are strongly affected by the superposition of isotropic
and anisotropic fluctuations \cite{abmp99}.
 Preliminary evidences  of the existence
of a scaling law also in the sectors with total angular
momentum $j=2$ have been reported \cite{arad_exp,sk00}. The value
of the exponent for the second order correlation function being
 close to the dimensional estimates $\xi_2^{j=2} = 4/3$,
\cite{lu67}.\\ All these preliminaries investigation in real
turbulent flows are flawed by the contemporary presence 
of anisotropies and strong
non-homogeneities. The very existence of scaling laws in presence
of strong non homogeneous effects can be doubted. SO(3) decomposition
becomes soon intractable as soon as non-homogeneous effects
cannot be neglected \cite{alp00}. 
Moreover, in many experimental situation, anisotropies
are introduced by a shear forcing  coupled to all turbulent scales: something
which prevents the possibility to study ``pure'' inertial physics. \\
To overcome this problems we performed  the first  numerical investigation
of a turbulent flow with strong anisotropic forcing  confined
to large scales and  {\it perfectly
homogeneous} on a numerical resolution $128^3$ and $256^3$.
 We studied  a fully periodic Kolmogorov
flow with random, delta correlated in time, forcing phases, which we decide
to call a ``Random-Kolmogorov Flow'' (RKF). \\
In this letter we present direct measurement of scaling exponents in sectors
up to total angular momentum $j=6$. Our main results  support the existence
of a hierarchical organization of exponents,  i.e. continuous
 increase of exponents
as a function of $j$. We also found a much stronger intermittency
in the anisotropic sectors than in the isotropic one.
 We conclude 
with a few comments  and  proposals
for further work in the field. 
\noindent
Let us begin to expose a few technical details on the simulations.
We performed a direct
numerical simulation of a fully periodic flow   with  anisotropic
large scales forcing. In details,
we have chosen a random forcing pointing only in one direction, 
the $z$ axis,
 with spatial dependency on the $\hat{x}$ direction
only on  two wavenumbers $\k_1=(1,0,0), \k_2=(2,0,0)$.  Namely: 
 $\f_i(\k_{1,2}) = \delta_{i,3} f_{1,2} 
\exp(i\theta_{1,2})$ where $f_1,f_2$ are two constant amplitude and 
$\theta_1,\theta_2$ are  two random phases, delta correlated in time. The
random phases allows for a homogeneous statistics also in the -otherwise-
non-homogeneous direction spanned by the two wavenumbers, i.e. we have
instantaneously a large scale non-homogeneity in the $x$ direction
which is averaged out by the time evolution thanks to the random
re-shuffling of the forcing phases. 
  We studied the RKF  at 
resolution $128^3$ and $256^3$,
we collected up to $200$ eddy turn over times for the smallest resolution
and up to $50$ eddy turn over times for the largest resolution. Such a long
averaging is necessary because as in any
strongly anisotropic flow  we observe
the formation of persistent large scale structures inducing strong
oscillation of the mean energy in time.

 In Fig. 1 we show, for example,
a typical time evolution for the total energy and total
energy dissipation in our runs, it is interesting to notice how
the high frequencies oscillation at  large scales (total energy)
induces by the random forcing are completely absent at small
scales (energy dissipation).\\  
In order to increase the scaling range extension we have used
an hyperviscosity with a squared laplacian.\
 Inset of Fig. 1  quantifies our degree of homogeneity.
 We have a high degree of homogeneity (more than $95\%$)
in the two transverse directions, $\hat{y},\hat{z}$, 
while we still observe small oscillations
in the $\hat{x}$ directions 
(of the order of $10 \%$); these oscillations are due to statistical
fluctuations, they must averaged out in the limit of infinite
statistics. \\ 
Let us now discuss the SO(3) decomposition of
longitudinal structure functions:
\begin{equation}
S_p(\r) = \la \left[(\v(\x)-\v(\x+\r))\cdot\unitr\right]^p \ra,
\end{equation}
where we have  kept only the dependency on $\r$ neglecting the
small non homogeneous fluctuations. 
We expect that the undecomposed structure functions are not the
``scaling'' bricks in the theory. Theoretical and numerical analysis
showed that one must first decompose the structure functions on the
irreducible representations of the rotational group and than asking
about the scaling behavior of the projection. In practice,
being the longitudinal structure functions scalar objects, 
their decompositions
reduces to the projections on the  spherical harmonics:
\begin{equation}
S_p(\r) = \sum_{j=0}^{\infty}\sum_{m=-j}^{j} S_p^{jm}(|r|) Y_{jm}(\unitr).
\label{so3_sf}
\end{equation} 
Where we have used the indeces $j,m$ 
to label the total angular momentum and its  projection
on a reference axis, say $\hat{z}$,  respectively. 
The whole physical information is hidden in the 
 coefficients $S_p^{jm}(|r|)$. In particular, the main 
question we want to address here concerns their  scaling properties:
$S_p^{jm}(|r|) \sim 
A_{jm} |r|^{\xi^j(p)}$ and (in the case) 
what one can say about the values of the scaling exponents, and 
on their robustness against large scale physics (universality
issue). Theoretical arguments suggest that if scaling exponents
exist they depend  only on the
$j$ eigenvalue \cite{foot}. If true turbulence follows
the Kraichnan models behavior, we should expect  universality
of the scaling exponents (independence of  large-scale boundaries),
  no saturation of the hierarchy 
($\xi^j(p) < \xi^{j'}(p)$ if $j < j'$) 
and strong non universalities in the prefactors $A_{jm}$. \\
We first present in Fig. 2 results concerning
the isotropic sector, $j=0,m=0$, comparing the undecomposed
structure functions in the three direction with the projection
$S_p^{00}(|r|)$ and their logarithmic local slopes (inset).
  Only for
the projected correlation it
 is possible to measure ($5\%$ of accuracy) the scaling
 exponents by a direct log-log fit versus the scale separation, $|\r|$. The
best fit gives $\xi^{j=0}(2) = 0.70 \pm 0.03$.
The undecomposed structure functions are overwhelmed by anisotropic
effects at all scales which spoil completely the scaling behavior. 

In Fig. 3 we present an overview of all sectors $j,m$ which have a
signal-to-noise ratio high enough to ensure stable results \cite{foot1}.
Sectors with odd $j$s are absent due to the parity symmetry of our
observable. We  measure anisotropic fluctuations up to
$j=6$. 
 We  notice from Fig. 3 
a clear foliation in terms of the $j$ index: sectors with the 
same $j$ but different $m$s behaves very similarly. 
In Table 1 we present a more quantitative analysis  by 
showing the results for the best power law fit 
for structure functions of orders $p=2,4$.
  The first result we want to   notice is
the absence of any saturation for the 
exponents as a function of the $j$
value. Unfortunatly the presence of an oscillation in all $j=2$ sectors
prevents us from measuring with accuracy the exponents in this sector, we
therefore refrain from giving  any number in this case. 

Let us also notice that the values for $j=4$ and $j=6$ are different
from what one would have expected if the anisotropic effects
would be given by simple smooth large scale fluctuations (see Table 1).
 This fact leads to the conclusion that anisotropies
are certainly the results of  non-linear interactions  in our flows, whether
they  corresponds to ``homogeneous'' fluctuations like
in the Kraichnan models  or 
to some dimensional balancing between the non-linear terms
and the forcing term is still an open question. The presence of a hierarchical
monotonic increasing of exponents at fixing $p$ and changing $j$ leads
to the possible breaking of the locality assumption 
in  high enough $j$ sectors \cite{loca}.
 For locality here we mean the fact that
all integrals of pressure-velocity correlation functions are convergent
both in the IR and in the UV limits.  \\
Let us conclude by assessing also the important point connected to the
existence of intermittency in higher $j$ sectors. From Table 1
we see that already for the $j=4$ sector, and even more for $j=6$,
the fourth order scaling exponents are ``almost'' saturated, i.e. very close
to the values of the second order exponents. It is hard to say
how much such a result is a quantitative 
sign of strong intermittency, due to the fact that
we lack a clear unambiguous dimensional --non intermittent-- prediction
for anisotropic exponents (see below).
 A fast saturation of exponents
within each sector as a function of the order of the moment must 
somehow be expected. 
  We imagine the statistics in the anisotropic
sectors being strongly dominated by ``persistent'' large
scale structures, introducing  cliffs
structures (statistically speaking) characteristic
also of saturation of exponents  in anisotropic scalar advection 
\cite{pass1,pass2}. Saturation,   in the anisotropic sectors
as a function of the order of the observed moment, $p$,
may also lead to the appearance of ``persistency of anisotropies''
even in presence of the observed  strict hierarchical ordering,
($\xi^j(p) < \xi^{j'}(p)$ if $j < j'$) as remarked  in \cite{bv00}.
 In conclusion we have presented the first numerical exploration
of an anisotropic homogeneous turbulent flow. We have confirmed
that by decomposing longitudinal structure functions in terms
of the eigenvectors of the rotational operator we have
a dramatic improvement of the scaling behavior in the isotropic
sector. We have also used
the SO(3) decomposition in order to asses two important questions
opened in the field  of anisotropic turbulence: (i)  the 
presence of a hierarchical organization of turbulent fluctuations
as a functions of the degree of anisotropy labeled by the $j$
index (ii) the existence of intermittency (saturation as a function
of the structure function order) in 
anisotropic sectors. 
The numerically measured values for the scaling exponents
$\xi^j(p)$ are  not consistent with a simple
``smooth'' hypothesis for the nature of anisotropic fluctuations.
More work is needed in order to understand the universality
degree of our results as a function of the anisotropic
properties of the large scale forcing. More work will be also
devoted to measure fully tensorial quantities like 
$D_{ij}(\r)= \la (v_i(\x)-v_i(\x+\r)) (v_j(\x)-v_j(\x+\r)) \ra$ 
in order to be  able to probe also odds sectors of the SO(3) group.\\

We conclude by noticing that dimensional prediction 
for the $\xi^j(p)$ with $j>0$ are far from being trivial. Indeed, different
dimensionless  quantities can be built by using some anisotropic mean
observable (the mean shear for example, or the mean squared shear in our
RKF) and the  energy dissipation.
Dimensional predictions than, would depend on the requirement that 
the anisotropic correction is (or is not) an analytical, smooth deviation
from  the isotropic sector. \\
The authors thank G. Boffetta and A. Celani for many 
helpful discussion and collaboration at the beginning of this work.
We also acknowledge useful discussions
with I. Arad, A. Lanotte, D. Lohse, V. L'vov and I. Procaccia. 
This research was supported in part by the EU under the Grant 
No. HPRN-CT  2000-00162 ``Non Ideal Turbulence'' and by the INFM  (Iniziativa
di Calcolo Parallelo).

\newpage

\begin{table}[htb]
\twocolumn[\hsize\textwidth\columnwidth\hsize\csname
@twocolumnfalse\endcsname
\begin{tabular}{||c|c|c|c|c|c||}
(j,m) & (0,0)  & (4,0) & (4,2)  &  (6,0) & (6,2) \\
\tableline
$\xi_2 | \xi^S_2$ & 0.70 $\pm$ 0.03 $|$ 2& 1.67 $\pm$ 0.07 $|$ 2& 1.7 $\pm 0.1$ $|$ 2 & 3.4 $\pm$ 0.2 $|$ 4  & 3.3 $\pm$ 0.2 $|$ 4 \\
$\xi_4 | \xi^S_4$ & 1.28 $\pm$ 0.05 $|$ 4& 2.15 $\pm$ 0.1 $|$ 4& 2.2 $\pm$ 0.1 $|$ 4 &  3.2 $\pm$ 0.2 $|$ 4 & 3.2 $\pm$ 0.2 $|$ 4 \\
\end{tabular}
\caption{Best fit of the scaling exponents in all 
stable sectors. For comparison we also give, $\xi_p^S$,
 the exponents for the case of a smooth (many times differentiable)
 anisotropic field. Some sectors are absent due either to the small 
signal-to-noise ratio or to the presence of sign changes 
in $S_p^{jm}(|\r|)$ which prevent the very definition of a slope. 
Errors are estimates from the fluctuation of the logarithmic local slopes at
resolution  $256^3$.}
]
\end{table}

\newpage
\begin{figure}
\hskip -.4cm\epsfig{file=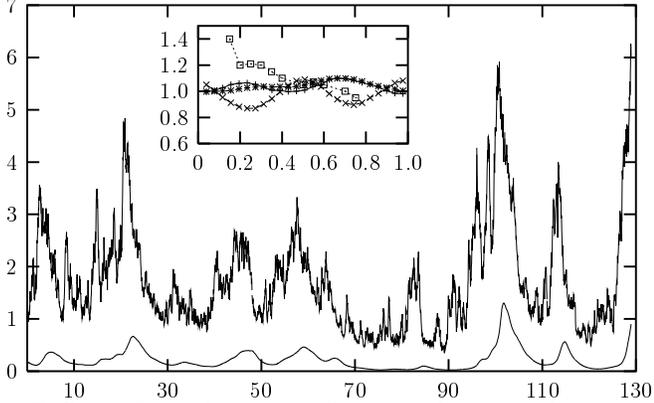,width=\hsize}
\caption{Typical  energy (above) and  energy dissipation (below)
time evolution in arbitrary units of the Random-Kolmogorov
flow at resolution $L_x=L_y=L_z=256$.
 Inset:  root mean squared
velocity $\la v^2_x \ra$ as a function of the spatial location in the
three directions: $\la v_x^2(x/L_x) \ra$ ($\times$), 
 $\la v_x^2(y/L_y) \ra$ ($\star$),
  $\la v_x^2(z/L_z) \ra$ ($+$).
For comparison is also shown the same quantity ($\Box$) 
from experimental state-of-the-art
 anisotropic  homogeneous shear flow at changing the 
position along the  shear direction $\hat{y}$
 [2]. All curves are normalized to be $1$ at $x/L_x=y/L_y=z/L_z=0.5.$}
\end{figure}
\newpage
\begin{figure}
\hskip -.4cm\epsfig{file=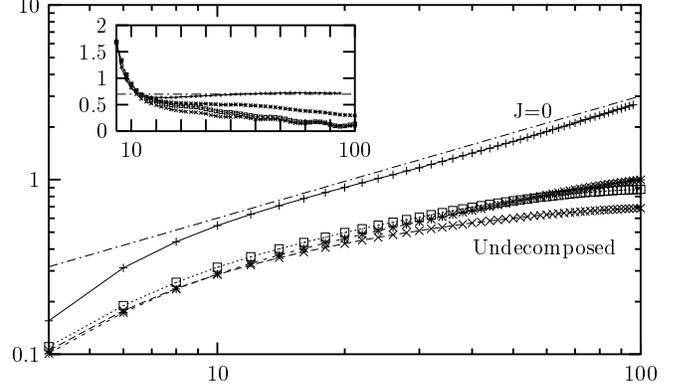,width=\hsize}
\caption{Isotropic sector. Log-log 
plot of  $S_2^{0,0}(|r|)$
versus $|r|$ ($+$), and the three undecomposed longitudinal structure functions
in the three directions $x,y,z$ ($\Box$,$\star$,$\times$) respectively,
at resolution $256^3$. The straight line has  the best fit 
slope $\xi^{j=0}(2)=0.70$.
 Inset: logarithmic
local slopes of all curves (same symbols) plus the straight line corresponding
to the intermittent isotropic scaling. 
Notice the dramatic improvement in the scaling
behavior of the projected correlation. Similar results hold for
higher orders $p>2$ (not shown).}
\end{figure}
\newpage
\begin{figure}
\hskip -.4cm\epsfig{file=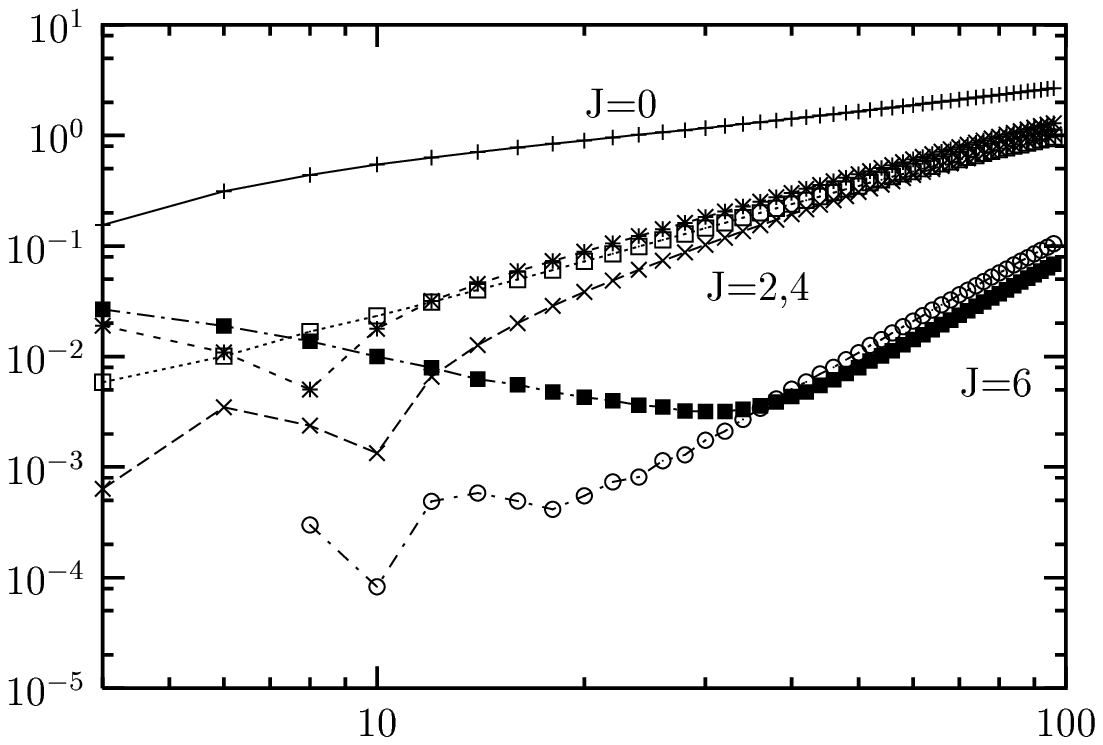,width=\hsize}
\caption{Log-log plot  for the absolute value of the projected
 second order structure functions, $|S_2^{j,m}(|r|)|$, versus 
the scale $r$,  on  all
measurable sectors (up to $j=6$). 
Sectors: $(0,0)$, ($+$); $(2,2)$, ($\times$);
$(4,0)$, ($\Box$); $(4,2)$, ($\star$); $(6,0)$, ($\circ$);
 $(6,2)$, ($\blacksquare$). The statistical
and numerical noise induced by the SO(3) projection 
can be estimate as the threshold where the $j=6$ sector
starts to deviate from the monotonic decreasing  behavior, i.e. 
$O(10^{-3})$.}
\end{figure}


\begin{thebibliography}{99}
\bibitem{foot} The $m$ eigenvalue 
does not play  any  role in fixing  the value
of the scaling exponents if they are determined by the homogeneous
scaling invariant part of the NS equations This is because the
non-linear inertial operator gouverning the correlation function
evolution depends only on the $j$ eigenvalue
(see \cite{alp00} for more details). 
The scaling exponents may depend on $m$ if they are
fixed by a matching with the forcing non-homogeneous term.
\bibitem{foot1} Some sectors in Fig. 3 are absent due to strong oscillation
in the projected structure functions $S_n^{jm}(|r|)$. These oscillations
are probably connected to strong large scale effects felt by those
particular spherical harmonics. We do not expect robustness in these cases,
different forcing and/or boundary conditions may remove the oscillatory behavior.
\end{thebibliography}
\end{document}